\documentclass[a4paper,11pt]{article}
\usepackage{jcappub} 
\usepackage{multirow}
\usepackage{color}    
\RequirePackage{graphicx}
\usepackage{amsmath}
\usepackage{graphicx}
\usepackage{subcaption}
\usepackage{tabularx}
\usepackage[a4paper, margin=2cm]{geometry}  

\title{\boldmath The Impact of 2D and 3D BAO Measurements on the {Cosmic} Distance Duality Relation with HII Galaxies}   

\author[a]{Jie Zheng,}
\author[a]{Da-Chun Qiang,}
\author[a]{Zhi-Qiang You,}
\author[a,1]{and Darshan Kumar\note{Corresponding author.}}
\affiliation[a]{Institute for Gravitational Wave Astronomy, Henan Academy of Sciences, Zhengzhou 450046, Henan, China}

\emailAdd{zhengjie@mail.bnu.edu.cn}  
\emailAdd{dcqiang@hnas.ac.cn}
\emailAdd{you\_zhiqiang@whu.edu.cn}
\emailAdd{kumardarshan@hnas.ac.cn }

\abstract{The cosmic distance duality relation (CDDR) is a fundamental and practical condition in observational cosmology that connects the luminosity distance and angular diameter distance. Testing its validity offers a powerful tool to probe new physics beyond the standard cosmological model. In this work, for the first time, we present a novel consistency test of CDDR by combining HII galaxy data with a comprehensive set of Baryon Acoustic Oscillations (BAO) measurements. The BAO measurements include two-dimensional (2D) BAO and three-dimensional (3D) BAO from the Sloan Digital Sky Survey (SDSS), as well as the latest 3D BAO data from the Dark Energy Spectroscopic Instrument (DESI) Data Release 2 (DR2). We adopt four different parameterizations of the distance duality relation parameter, $\eta(z)$, to investigate possible deviations and their evolution with cosmic time. To ensure accurate redshift matching across datasets, we reconstruct the distance measures through a model-independent Artificial Neural Network (ANN) approach. {We find no significant deviation from the CDDR (less than 68\% confidence level) among four parameterizations. Furthermore, our results show that the constraints on $\eta(z)$ obtained separately from 2D and 3D BAO measurements are consistent at the 68\% confidence level. This indicates that there is no significant tension between the two datasets under the four parameterizations considered. Our ANN reconstruction of HII galaxies could provide constraints on the CDDR at redshifts beyond the reach of Type Ia supernovae.} Finally, the consistency of our results supports the standard CDDR and demonstrates the robustness of our analytical approach.

\vspace{2mm}
}   

\begin{document}
\maketitle
\flushbottom

\section{Introduction}\label{sec:intro}

The cosmic distance duality relation (CDDR), also known as the Etherington relation \cite{1933PMag...15..761E}, is a cornerstone of modern observational cosmology. It provides a direct connection between the luminosity distance ($d_{L}$) and the angular diameter distance ($d_{A}$) through the expression $d_{L}(z) = d_{A}(z)(1+z)^{2}$, where $z$ denotes the redshift. This relation, first proposed by Etherington, relies on three fundamental assumptions \cite{2007GReGr..39.1047E}: $i)$ spacetime is described by a metric theory of gravity, $ii)$ photons travel along unique null geodesics, and $iii)$ photon number is conserved. Under these conditions, the CDDR is expected to hold at all redshifts within the standard cosmological framework, corresponding to a theoretical prediction of $\eta(z) = d_{L}(z)/\left(d_{A}(z)(1+z)^{2}\right) = 1$. Therefore, any significant deviation from the standard CDDR may indicate the breakdown of one or more of these assumptions, pointing toward possible new physics, such as photon–axion conversions \cite{2004PhRvD..69j1305B,2004ApJ...607..661B}, cosmic opacity induced by intergalactic dust or exotic interactions \cite{2009ApJ...696.1727M,2012JCAP...12..028N}, and modifications to general relativity \cite{2004PhRvD..70h3533U,2017PhRvD..95f1501S,2021PhRvD.104h4079A}.


The increasing tensions between cosmological parameters inferred from early- and late-Universe observations \cite{Riess2022,1807.06209,2012MNRAS.427..146H,2020NatAs...4..196D,2021PhRvD.103d1301H,2021A&A...646A.140H,2021ApJ...908L...9D,2022PhRvD.105b3520A} have raised growing concerns about the robustness of the standard cosmology. The most significant one is so-called the ``Hubble tension", which has been observed by various cosmological probes \cite{2018AAS...23231902P,2020MNRAS.498.1420W}, reaching a statistical significance exceeding 5$\sigma$ and indicating either the existence of new physics beyond the standard model or the need to re-examine some of its foundational assumptions \cite{2016PhLB..761..242D,pantheon,2020arXiv200710716E,Riess2022}. One of these assumptions is the validity of the CDDR. Recent studies \cite{Renzi2023,Li2023,2025arXiv250410464T,2025arXiv250115233A} suggest that a violation of the CDDR could introduce inconsistencies in the calibration of distance measurements from different cosmological probes, potentially contributing to the observed tensions. Therefore, it is essential to test the validity of the CDDR in the context of the current cosmological tensions.

{A considerable amount of research has focused on testing the CDDR employing a variety of cosmological observations, which requires simultaneous measurements of luminosity distances ($d_{L}(z)$) and angular diameter distances ($d_{A}(z)$). Type Ia supernovae (SNIa) are commonly used to determine luminosity distances, while angular diameter distances have been provided by different cosmological probes, such as the Sunyaev–Zeldovich effect and gas mass fraction measurements in galaxy clusters \cite{2010ApJ...722L.233H,2011ApJ...729L..14L,2013MNRAS.436.1017L}, baryon acoustic oscillations (BAO) \cite{Wu:2015ixa,2025EPJC...85..186Y,2020EPJP..135..447X}, strong gravitational lensing systems \cite{2017JCAP...03..028R,2017JCAP...07..010R,2021PhRvD.103f3511K,2022JCAP...01..053K,2011JCAP...05..023N,2021ChPhC..45a5109L,2025ApJ...979....2Q}, and the angular size of ultra-compact radio sources \cite{2018MNRAS.474..313L,2022ChJPh..78..297H}. The combination of these diverse probes has enabled multifaceted examinations of the CDDR across different cosmic epochs and distance scales \cite{2010JCAP...10..024A,2010JCAP...02..008S,2012JCAP...06..022H,2013JCAP...04..027H,2013PhLB..718.1166L,2017IJMPD..2650097F,2025arXiv250711518K}.}

{Despite these numerous studies, several recent developments make a renewed investigation both timely and necessary. First, the Dark Energy Spectroscopic Instrument (DESI) collaboration has released its second dataset (DR2, hereafter 3D-DESI), including observations from the first three years of operation \cite{2025JCAP...04..012A}. This release delivers the most precise BAO measurements to date. However, analyses of this dataset have reported potential tensions with the $\Lambda$CDM model \cite{2025JCAP...04..012A,2025JCAP...02..021A,2024arXiv241204830Z,2024PDU....4601668H,2025ApJ...987...58W,2025arXiv250609819L,2025arXiv250502207Y,2025arXiv250509470A,2025arXiv250524732C}, prompting further investigation into the underlying assumptions of standard cosmology. Since the CDDR relies on these assumptions, testing its validity with the latest BAO measurements may shed light on the origin of the deviations. Moreover, several studies have reported the disagreements between BAO measurements obtained from the two-dimensional (2D, transverse or angular) BAO and the three-dimensional (3D, or anisotropic) BAO \cite{2019PhRvD..99l3515A,2020MNRAS.495.2630C,2023PhRvD.107j3531B,9rt7-ph33,2024Univ...10..406D,2024PhLB..85839027F}, raising the possibility that CDDR tests based on different BAO types may yield different results. Recently, \cite{2024PhLB..85839027F} investigated how such tensions could affect the CDDR by performing a model-independent analysis that used SNIa to provide the luminosity distances, finding no significant violation of the relation. In this work, we revisit this issue by combining the latest HII galaxy data with different BAO datasets, and perform a parameterized analysis with four different forms of $\eta(z)$. This strategy enables us to probe potential deviations from the CDDR while directly assessing how the 2D and 3D BAO tension might affect the inferred behavior of $\eta(z)$. }

{We choose HII galaxy data as the luminosity distance probe in this analysis for several reasons. First, while SNIa are the most commonly used standard candles, an important consideration is their dependence on the absolute peak magnitude $M_{B}$, which is traditionally assumed to be constant. However, recent studies suggest that $M_{B}$ may evolve with redshift \cite{2022JCAP...01..053K,2020MNRAS.495.2630C,2023PhRvD.107f3513D,2023PDU....3901160B,2025arXiv250415127V,2021MNRAS.504.5164C,2020PhRvR...2a3028C}, potentially introducing additional uncertainties in the luminosity distance measurements and affecting the robustness of CDDR tests \cite{2023PDU....3901160B,2025arXiv250415127V,2021MNRAS.501.3421K,2020PhRvD.102b3520K}. By contrast, HII galaxy data and giant extragalactic HII regions (GEHR) data provide a viable alternative. Their redshift coverage extends up to $z \sim 2.5$, overlapping well with BAO datasets. In addition, they exhibit a robust correlation between the H$\beta$ luminosity $L(\mathrm{H}\beta)$ and the ionized gas velocity dispersion $\sigma$, enabling an independent determination of luminosity distances. Furthermore, their sensitivity to photon-number nonconservation makes them particularly well-suited for model-independent CDDR tests.}

{One of the main challenges in testing the CDDR is obtaining matched luminosity and angular diameter distances at the same redshifts. In our case, we reconstruct the luminosity distance–redshift relation from HII galaxy data using the Artificial Neural Networks (ANN) method. Compared to traditional non-parametric techniques, the ANN method offers several distinct advantages. It is fully data-driven, imposes no assumptions on the statistical distribution or functional form of the underlying relationship, and can flexibly capture complex non-linear patterns in the data. Moreover, as a universal approximator, the ANN method can accurately model any continuous function, provided that the hidden layer contains a sufficient number of neurons. This approach has been demonstrated to be effective in cosmological research \cite{2020ApJS..246...13W,2022JCAP...02..023D,2023PDU....3901160B,2023PhRvD.108f3522Q,2023ChPhC..47a5101T,2024PDU....4601706M,2024EPJC...84....3Y,2025MNRAS.540.2253A,2025ApJ...987...58W,2025JHEAp..4700377H,2025ApJS..276...71L,2025ApJ...979....2Q}. Hence, the ANN method is particularly suitable for our work, since its strengths allow for a model-independent and robust reconstruction of luminosity distances.}

In this work, we perform the first systematic CDDR test that jointly uses HII galaxy data and various BAO datasets. This approach not only enables a robust parameterized determination of $\eta(z)$, but also allows us to assess whether the tensions between different BAO measurements would result in the deviation of $\eta(z)$. The outline of the paper is as follows: In section~\ref{sec:data}, we discuss the Data and Methodology. The analysis and results are explained in section~\ref{sec:discuss}. Finally, the discussions and conclusions are presented in section~\ref{sec:conclusion}.

\section{Data and Methodology}\label{sec:data}
In this section, we present the details of the observational datasets (BAO and HII galaxies) and our methodology adopted for CDDR validation. 
\subsection{The BAO datasets}
The clustering of matter imprinted by BAO serves as a ``standard ruler'' in cosmology, with its length set by the sound horizon at the drag epoch, denoted as $r_d$. During the drag epoch, baryons decoupled from photons and the BAO scale was ``frozen in'' at the sound horizon, $r_d = r_s(z_d)$, where $z_d$ is the redshift of the drag epoch. The sound horizon is given by
\begin{equation}
\label{eq:rdequation}
r_d = \int_{z_d}^\infty \frac{c_s(z)}{H(z)}dz,
\end{equation}
where $c_s(z)$ is the sound speed and $H(z)$ is the Hubble parameter. When using BAO measurements for cosmological studies, it is crucial to know the length of this standard ruler, as it enables the exploration of dark energy and the Universe’s expansion history.

Galaxy surveys have succeeded in determining the angular BAO scale, $\theta_{\mathrm{BAO}}$, defined by
\begin{equation}
\label{eq:theta}
\theta_{\mathrm{BAO}} = \frac{r_d}{(1+z)d_A(z)},
\end{equation}
where $d_A(z)$ is the angular diameter distance and the comoving distance is $d_M(z) = (1+z)d_A(z)$. In this work, we use two types of BAO datasets: the angular (2D) BAO data, consisting of 15 measurements of $\theta_{\mathrm{BAO}}$ at various redshifts (see Table.~\ref{tab:2D}), and the anisotropic (3D) BAO data, presented as $d_A(z)/r_d$ (see Table.~\ref{tab:3D}). The 2D-BAO data are derived from SDSS data releases DR7, DR10, DR11, DR12, and DR12Q \cite{2020MNRAS.497.2133N,2021A&A...649A..20D,2016PhRvD..93b3530C,2016arXiv161108458A,2020APh...11902432C,2018JCAP...04..064D}, obtained without assuming a fiducial cosmological model. For the 3D-BAO analysis, we consider two datasets: one from DES Y6 and BOSS/eBOSS \cite{2017MNRAS.465.1757G,DES:2024pwq,2021MNRAS.500.1201H,2020ApJ...901..153D,2017MNRAS.470.2617A,2021PhRvD.103h3533A}, and another from recent DESI DR2 results \cite{2025JCAP...04..012A}. To ensure model and calibrator independence, only the angular components of the 3D BAO measurements are used, with radial and dilation scale data excluded.

\begin{figure}[htbp]
    \centering
    \includegraphics[width=0.7\linewidth]{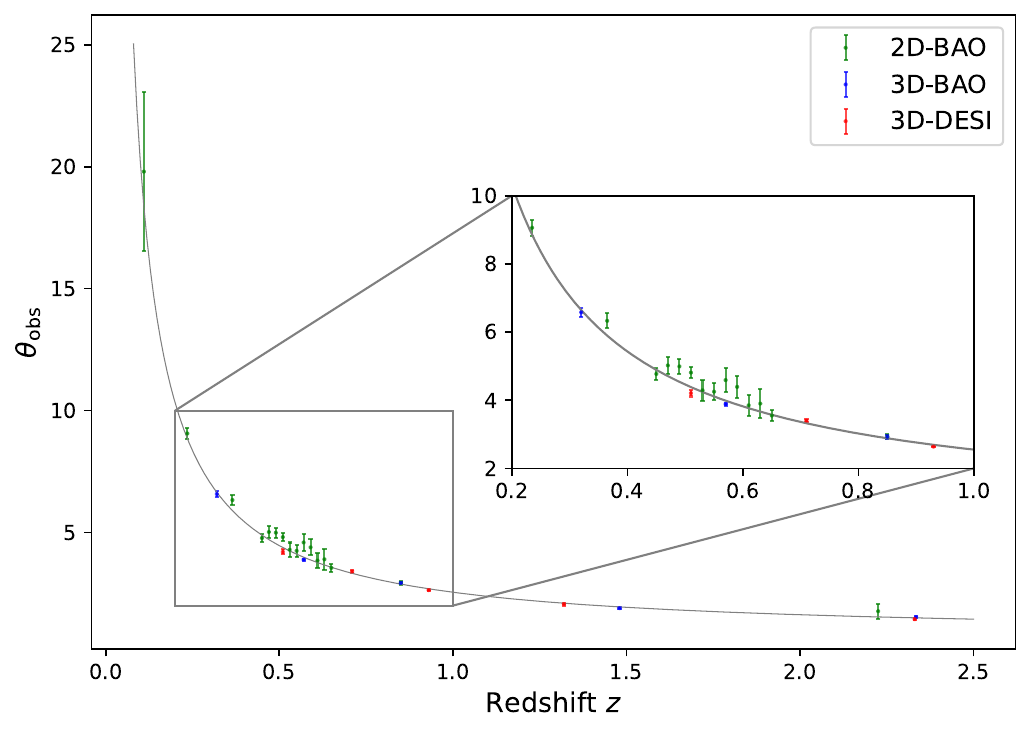}
    \caption{The 2D-BAO, 3D-BAO and 3D-DESI measurements of $\theta(z)=r_d/d_{M}(z)$. The grey line corresponds to the theoretical values of $\theta(z)$ from the $\Lambda$CDM model with $\Omega_{m} = 0.3$ and $H_0 = 70$ km/s/Mpc}
    \label{fig:theta_obs}
\end{figure}


\begin{table}[htbp]
\renewcommand{\arraystretch}{1.4}                                          

\centering
\begin{tabular}{|l|c|c|p{4.5cm}|}
\hline
Survey & $z$ & $\theta_{\mathrm{BAO}}$ [deg] & References \\
\hline
SDSS DR12 & 0.11 & $19.8 \pm 3.26$ & de Carvalho et al. (2021) \\
\hline
\multirow{2}{*}{SDSS DR7} 
          & 0.235 & $9.06 \pm 0.23$ & \multirow{2}{*}{Alcaniz et al. (2017)} \\ 
\cline{2-3}
          & 0.365 & $6.33 \pm 0.22$ & \\
\hline
\multirow{6}{*}{SDSS DR10} 
          & 0.45  & $4.77 \pm 0.17$ & \multirow{6}{*}{Carvalho et al. (2016)} \\
\cline{2-3}
          & 0.47  & $5.02 \pm 0.25$ & \\
\cline{2-3}
          & 0.49  & $4.99 \pm 0.21$ & \\
\cline{2-3}
          & 0.51  & $4.81 \pm 0.17$ & \\
\cline{2-3}
          & 0.53  & $4.29 \pm 0.30$ & \\
\cline{2-3}
          & 0.55  & $4.25 \pm 0.25$ & \\
\hline
\multirow{5}{*}{SDSS DR11} 
          & 0.57  & $4.59 \pm 0.36$ & \multirow{5}{*}{Carvalho et al. (2020)} \\
\cline{2-3}
          & 0.59  & $4.39 \pm 0.33$ & \\
\cline{2-3}
          & 0.61  & $3.85 \pm 0.31$ & \\
\cline{2-3}
          & 0.63  & $3.90 \pm 0.43$ & \\
\cline{2-3}
          & 0.65  & $3.55 \pm 0.16$ & \\
\hline
BOSS DR12Q & 2.225 & $1.77 \pm 0.31$ & de Carvalho et al. (2018) \\
\hline
\end{tabular}
\caption{List of the 15 2D BAO data points used in this work, with $\theta_{\mathrm{BAO}}(z)\,[\mathrm{rad}] = r_d / [(1 + z)d_A(z)]$. The values in the third column are given in degrees. See the quoted references for details.}
\label{tab:2D}
\end{table}

\begin{table}[htbp]
\renewcommand{\arraystretch}{1.4}

\centering
\begin{tabular}{|l|c|c|p{5.9cm}|}
\hline
Survey & $z$ & $d_A(z)/r_d$ & References \\
\hline
\multirow{2}{*}{BOSS DR12}
        & 0.32 & $6.5986 \pm 0.1337$ & \multirow{2}{*}{Gil-Mar\'in et al.\ (2017)} \\
\cline{2-3}
        & 0.57 & $9.389 \pm 0.103$ & \\
\hline
DES Y6 & 0.85 & $2.932 \pm 0.068$ & Abbott et al. (2024a) \\
\hline
eBOSS DR16Q & 1.48 & $12.18 \pm 0.32$ & Hou et al.\ (2020) \\
\hline
eBOSS DR16 Ly$\alpha$-F & 2.334 & $11.25^{+0.36}_{-0.33}$ & du Mas des Bourboux et al.\ (2020) \\
\hline
{DESI DR2 LRG1} & 0.510  & $8.998 \pm 0.112$  & \multirow{6}{*}{Abdul Karim et al. (2021)} \\
\cline{1-3}
   {DESI DR2 LRG2}     & 0.706  & $10.168 \pm 0.106$ & \\
\cline{1-3}
   {DESI DR2 LRG3+ELG1}     & 0.934  & $11.155 \pm 0.080$ & \\
\cline{1-3}
   {DESI DR2 ELG2}     & 1.321  & $11.894 \pm 0.138$ & \\
\cline{1-3}
   {DESI DR2 QSO}     & 1.484  & $12.286 \pm 0.305$ & \\
\cline{1-3}
    {DESI DR2 Ly$\alpha$}    & 2.330  & $11.708 \pm 0.159$ & \\
\hline
\end{tabular}
\caption{Summary of 3D BAO measurements used in this work, with $d_A(z)/r_d$. See the quoted references for details. As explained in section~\ref{sec:data}, we employ two alternative 3D BAO datasets: the first five rows correspond to the BOSS/eBOSS data points, while the remaining rows include measurements from the DESI DR2.} 
\label{tab:3D}
\end{table}               


{As BAO measurements fundamentally rely on the value of $r_d$, the adopted treatment of $r_d$ can affect both the estimated angular diameter distances $d_A(z)$ and the subsequent reconstruction of the $\eta(z)$ function. To avoid making any assumption, we treat $r_d$ as a nuisance parameter in our analysis {and numerically marginalize over} it with broad, non-informative priors. Unless otherwise stated, all constraints reported below are marginalized over $r_d$.}

\subsection{The HII galaxy sample}

To estimate the luminosity distance, we analyze a full sample of 181 HII galaxies (HIIGx) in the redshift range $0.01 < z < 2.6$ \cite{2021MNRAS.505.1441G}. The sample consists of 74 high-redshift HIIGx observed in the range $0.5 < z < 2.6$ \cite{2019MNRAS.487.4669G}, and 107 local HIIGx with redshifts in the interval $0.01 < z < 0.2$ \cite{2014MNRAS.442.3565C}. 

HIIGx are compact systems undergoing intense starburst episodes that dominate their total luminosity. Their optical spectra are characterized by strong Balmer emission lines, especially H$\alpha$ and H$\beta$, which result from the recombination of hydrogen ionized by young, massive stellar populations \cite{1987MNRAS.226..849M,1972ApJ...173...25S,1977ApJ...211...62B,1981MNRAS.195..839T}. These systems share physical properties with giant extragalactic HII regions (GEHR), although GEHR are typically located in the outer disks of late-type spiral galaxies. A strong empirical correlation has been established between the H$\beta$ luminosity, $L(H\beta)$, and the velocity dispersion of the ionized gas, $\sigma(H\beta)$. This correlation, known as the $L$–$\sigma$ relation, shows a small intrinsic scatter and enables the use of HIIGx and GEHR as standard candles in cosmological analyses \cite{2000MNRAS.311..629M,2024EPJC...84....3Y,2005MNRAS.356.1117S}. 

The $L$–$\sigma$ correlation \cite{2012MNRAS.425L..56C,2014MNRAS.442.3565C} is given by
\begin{equation} \label{equ_hii_corr}
    \log_{10} \left[\frac{L(H \beta)}{\mathrm{erg}~\mathrm{s}^{-1}}\right]
    = \alpha \log_{10} \left[\frac{\sigma(H \beta)}{\mathrm{km}~\mathrm{s}^{-1}}\right] + \beta,
\end{equation}
where $\alpha$ and $\beta$ are empirical constants representing the slope and intercept of the relation, respectively. Using the definition of luminosity distance, the corresponding expression for the distance modulus becomes
\begin{equation} \label{equ_hii_dm}
    \mu = 5 \log_{10} \left(\frac{d_L}{\mathrm{Mpc}}\right) + 25 
    = 2.5\left[\alpha \log_{10} \left(\frac{\sigma(H \beta)}{\mathrm{km}~\mathrm{s}^{-1}}\right) 
    - \log_{10} \left(\frac{F(H \beta)}{\mathrm{erg}~\mathrm{s}^{-1}~\mathrm{cm}^{-2}}\right) + \beta\right] - 100.2,
\end{equation}
where $d_L$ is the luminosity distance and $F(H\beta)$ is the observed flux in the H$\beta$ emission line. 

Although the parameters $\alpha$ and $\beta$ are, in principle, nuisance parameters that should be fit jointly with cosmological parameters to avoid circularity, studies have shown that they are largely insensitive to the choice of cosmology. Therefore, we adopt the values $\alpha = 33.268 \pm 0.083$ and $\beta = 5.022 \pm 0.058$, as obtained in previous analyses \cite{2021MNRAS.505.1441G,2019MNRAS.487.4669G,2024EPJC...84....3Y,2023PhRvD.107j3521C}. 

The corresponding uncertainty in the distance modulus derived from Eq.~\ref{equ_hii_dm} is given by
\begin{equation}
    \sigma^2_{\mu} = 6.25\left(\sigma_{\log_{10} F}^2 
    + \beta^2 \sigma_{\log_{10} \sigma}^2 
    + \sigma_\beta^2 \left(\log_{10} \sigma\right)^2 
    + \sigma_\alpha^2\right),
\end{equation}
where $\sigma_{\log F}$ and $\sigma_{\log \sigma}$ represent the uncertainties in the logarithmic flux and velocity dispersion, respectively, and $\sigma_\alpha$ and $\sigma_\beta$ are the uncertainties associated with the fitted parameters.

\begin{figure}
    \centering
    \includegraphics[width=0.7\linewidth]{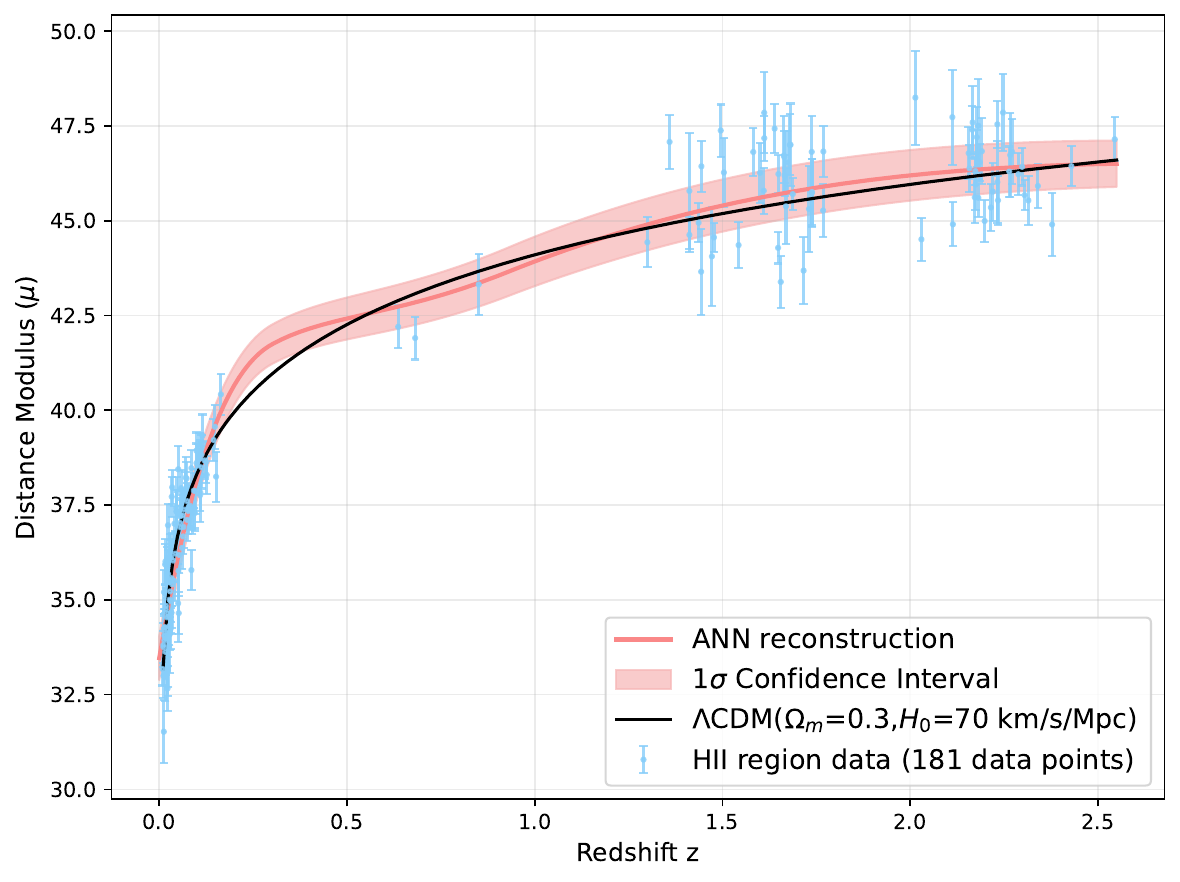}
    \caption{The observed values of $\log d_{L,\mathrm{HII}}(z)$ from HII galaxy measurements are shown as cyan data points, with error bars representing the 68\% confidence level. The red line represents the reconstructed function, and the pink shaded region indicates the 68\% confidence level obtained using the ANN method. The black line corresponds to the theoretical prediction of $\log d_{L,\mathrm{HII}}(z)$ from the $\Lambda$CDM model with $\Omega_{m} = 0.3$ and $H_0 = 70$ km/s/Mpc}
    \label{fig:HII} 
\end{figure}

{
We note that recent studies suggest the $L$–$\sigma$ relation is not sufficiently precise for use as a cosmological distance indicator at the highest redshifts. For example, Ref.~\cite{2024A&A...690A.157M} analyzed HII galaxies up to $z \sim 7$ with Giant Extragalactic HII Regions (GEHRs) and JWST data. Their work, along with related studies, reports that high-redshift samples show significantly larger scatter in the $L$–$\sigma$ relation. To avoid these less-standardized regimes, where systematics become more severe, we impose a redshift cutoff. Since the DESI DR2 dataset extends only to $z < 2.4$, we restrict our HII galaxy sample to the same range. This excludes the GEHR and JWST high-$z$ data, which contribute additional scatter and deviations in the relation. Within this redshift interval, HII galaxies have been consistently used as standard candles in cosmological analyses, with their reliability supported by multiple independent studies \cite{2024EPJC...84....3Y,2024arXiv240810560G,2025arXiv250613812R,2024PDU....4601641S}.

}

\subsection{Reconstruction Method: Artificial Neural Network}  

{The ANN method provides a fully data-driven and non-parametric framework for reconstructing functions from observational data. In particular, it avoids imposing prior assumptions on the cosmological model or the statistical distribution of the measurements \cite{2020ApJS..246...13W,2022JCAP...02..023D,2023PDU....3901160B,2023PhRvD.108f3522Q,2023ChPhC..47a5101T,2024PDU....4601706M,2024EPJC...84....3Y,2025MNRAS.540.2253A,2025ApJ...987...58W,2025JHEAp..4700377H,2025ApJS..276...71L,2025ApJ...979....2Q}. This flexibility allows ANNs to capture complex, non-linear relationships in observational datasets, even when the noise properties deviate from Gaussianity. Furthermore, as universal approximators, the ANN method can represent a wide range of functions to arbitrary precision given sufficient neurons and a suitable network design. These characteristics make ANNs particularly suitable for reconstructing the distance modulus $\mu(z)$ from HII galaxy data, for which we use the \texttt{REFANN} Python package \cite{2020ApJS..246...13W}.}

{The ANN is designed to learn a nonlinear mapping between the input and output based solely on observational measurements. In our case, the input variable is the redshift $z$, and the output to be learned is the distance modulus $\mu(z)$. The adopted network consists of one input neuron (for $z$), a single hidden layer with 4096 neurons, and one output neuron corresponding to the predicted $\mu(z)$ and its uncertainty $\sigma_{\mu}$. Each neuron performs a linear transformation followed by a non-linear activation, where we use the Exponential Linear Unit (ELU) \cite{2015arXiv151107289C}:}
\begin{equation}
f(x) =
\begin{cases}
x, & x > 0 \\
\alpha(e^x - 1), & x \leq 0
\end{cases},
\quad \alpha = 1.
\end{equation}

{The training process minimizes the least absolute deviation loss \cite{2020ApJS..246...13W}, which is more robust to non-Gaussian errors and outliers than mean-squared error, thus keeping our implementation consistent with \texttt{REFANN}:}
\begin{equation}
L = \frac{1}{mp} | \hat{Y} - Y |,
\end{equation}
{where $m$ is the batch size, $p$ is the number of output parameters, $\hat{Y}$ denotes the network's prediction, and $Y$ is the corresponding ground-truth values. Following the standard training configuration of \texttt{REFANN}~\cite{2020ApJS..246...13W}, we use gradient descent with the Adam optimizer (initial learning rate $0.01$) and a batch size equal to 50\% of the dataset. The network is trained over 30,000 iterations to ensure convergence for the HII galaxy $\mu(z)$ reconstruction. For each reconstruction, a single realization of the observational dataset is used for training, and the uncertainty in the reconstructed $\mu(z)$ is estimated by propagating the measurement errors through the trained network, following the standard \texttt{REFANN} procedure.}

{By training the ANN on the HII galaxy data, we obtain a smooth, model-independent reconstruction of $\mu(z)$ that captures the nonlinear relation between redshift and distance modulus (see in figure~\ref{fig:HII}). This allows us to directly derive the luminosity distance $d_L(z)$ at the BAO redshifts. Hence, we can make a model-independent comparison with the angular diameter distances $d_A(z)$ from BAO measurements and test the CDDR. While the ANN approach provides flexibility and robustness, its performance can be influenced by hyperparameter choices such as network architecture and training methodology. These factors may affect the flexibility and generalization of the model, and thus the reconstructed $\mu(z)$ and its confidence intervals. }

\subsection{Parameterizations of CDDR}
To explore the possibility of violation of the standard CDDR, we rewrite the relationship between angular diameter distance $d_A(z)$ and luminosity distance $d_{L}(z)$ at redshift $z$ as
\begin{equation}
\label{equ_cddr_defi}
    \eta(z)=\dfrac{d_L(z)}{\left(1+z\right)^2d_A(z)},
\end{equation}
where $\eta(z) = 1$ holds if the standard relation is valid, and any deviation of $\eta(z)$ from unity implies the violation of the CDDR. 
In this work, we examine four parameterizations of $\eta(z)$ \cite{2010JCAP...10..024A,2020JCAP...12..019H,2021PDU....3200824R,2022JCAP...01..053K,2025RAA....25b5019G,2025ChPhC..49a5104T,2025arXiv250410464T}, namely:

\begin{itemize}
    \item A linear parameterization, P1: $\eta(z) = 1 + \eta_1 z$,
    \item A modified linear parameterization, P2: $\eta(z) = 1 + \eta_1 \frac{z}{1+z}$,
    \item A logarithmic parameterization, P3: $\eta(z) = 1 + \eta_1 \ln(1+z)$,
    \item A power-law parameterization, P4: $\eta(z) = (1+z)^{\eta_1}$,
\end{itemize}
where the parameter $\eta_{1}=0$ corresponds to the standard CDDR. \\

{
In our analysis, we focus on two parameters: the cosmic distance duality relation $(\eta_1)$ and the sound horizon $(r_d)$. For $\eta_1$, we adopted a wide flat prior $\mathcal{U}[-2,2]$ to minimize prior-driven effects on the posterior distribution. Further, we performed marginalization over $r_d$ by integrating the posterior with a uniform prior $\mathcal{U}[10,300]$. This approach allows us to constrain $\eta_1$ robustly without imposing strong assumptions on the sound horizon. To perform the parameter estimation, we use Bayesian inference with the Python module \textbf{\texttt{emcee}}\footnote{\url{https://emcee.readthedocs.io/en/stable/}}, an affine-invariant Markov chain Monte Carlo (MCMC) sampler \cite{emcee}. The MCMC analysis uses 40 walkers and 40,000 steps for each walker to provide a thorough exploration of the parameter space. The initial 20\% of the samples from each chain are discarded as burn-in to eliminate potential biases from initial conditions. The remaining samples are used to construct the posterior distributions. We check the convergence of the chains in two ways. First, we visually inspected the trace plots for each parameter to confirm proper mixing and stability around the best-fit values. Second, we calculated the integrated auto-correlation time $\tau_f$ using the \textbf{\textit{autocorr.integrated\_time}} function from the \textbf{\textit{emcee}} package. This procedure ensures that the correlated nature of ensemble samplers and provides statistically consistent estimates of the posteriors.
}

\section{Results}
\label{sec:discuss}

\begin{figure}
    \centering
    \includegraphics[width=0.7\linewidth]{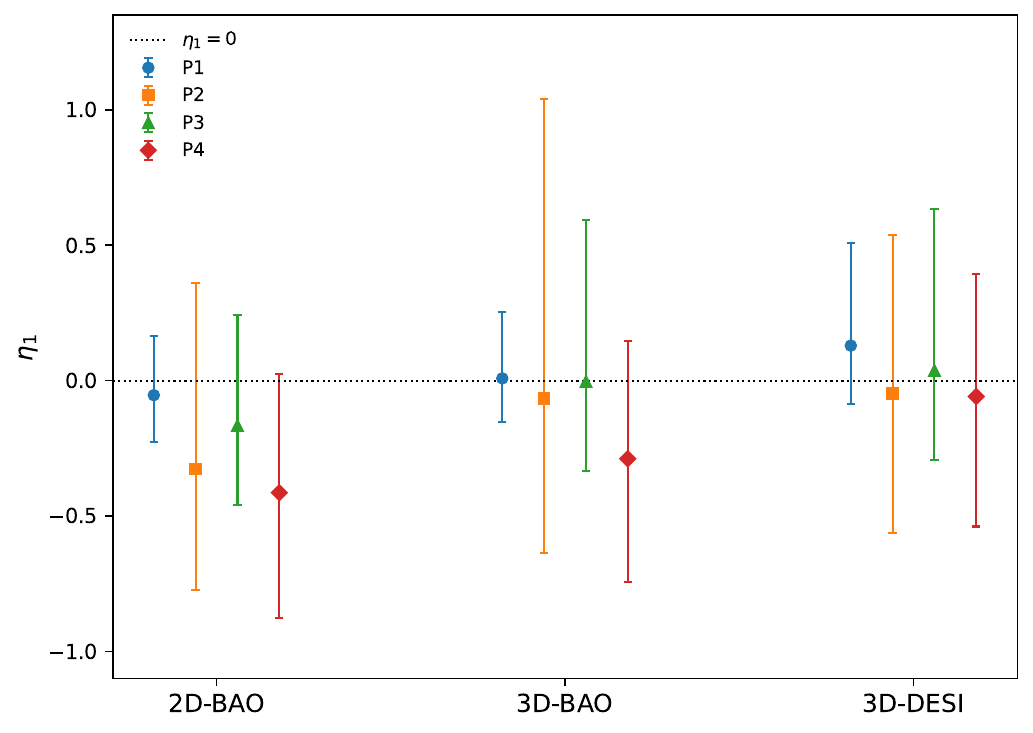}
    \caption{The normalized posterior distributions of $\eta$ obtained from HII galaxy data combined with 2D-BAO, 3D-BAO, and 3D-DESI BAO datasets. The vertical dashed line at $\eta = 0$ marks the value predicted by the standard CDDR.}
    \label{fig:result} 
\end{figure}

{We present the first test of the CDDR with HII galaxy data serving as luminosity distance indicators, providing an alternative to SNIa for measuring $d_L(z)$. Moreover, we consider three types of BAO datasets, 2D-BAO, 3D-BAO, and 3D-DESI, to quantify whether the tension between 2D angular and 3D anisotropic BAO datasets would affect the constraints on the parameterized CDDR, $\eta(z)$. Here, {we show the constraints obtained on $\eta_1$ with four different parameterizations}: $\eta(z)=1+\eta_1 z$ (P1), $\eta(z)=1+\eta_1 \frac{z}{1+z}$ (P2), $\eta(z)=1+\eta_1 \ln(1+z)$ (P3), and $\eta(z)=(1+z)^{\eta_1}$ (P4), as summarized in table~\ref{tab:results} and shown in figure~\ref{fig:result}. In particular, {we treat $r_d$ as a nuisance parameter, sampling from a prior uniform in $[10,~300]$  and marginalizing over it.}
}

{For the \textbf{P1} model, defined as $\eta(z) = 1 + \eta_1 z$, our analysis yields the following constraints: $\eta_{1}=-0.054^{+0.220}_{-0.172}$ from 2D-BAO, $\eta_1=0.008^{+0.246}_{-0.162}$ from 3D-BAO, and $\eta_{1}=0.129^{+0.379}_{-0.214}$ from 3D-DESI.
The constraints on $\eta_{1}$ are consistent with zero at the 68\% confidence level, supporting the validity of the CDDR.
In the case of the \textbf{P2} model, where $\eta(z) = 1 + \eta_1 \frac{z}{1+z}$, we obtain the following constraints: $\eta_{1}=-0.327^{+0.686}_{-0.448}$ from 2D-BAO, $\eta_1=-0.066^{+1.106}_{-0.572}$ from 3D-BAO, and $\eta_{1}=-0.047^{+0.584}_{-0.516}$ from 3D-DESI.
Similar to the P1 model, the results show no strong evidence for a violation of the CDDR. 
For the \textbf{P3} model, where $\eta(z) = 1 + \eta_1 \ln(1+z)$, our analysis yields the following constraints: $\eta_{1}=-0.166^{+0.409}_{-0.293}$ from 2D-BAO, $\eta_1=-0.004^{+0.598}_{-0.331}$ from 3D-BAO, and $\eta_{1}=0.037^{+0.344}_{-0.308}$ from 3D-DESI. 
These results yield $\eta_1$ values consistent with zero within a 68\% confidence level.
Besides, for the \textbf{P4} model, where $\eta(z) = (1+z)^{\eta_1}$, we obtain the following constraints: $\eta_{1}=-0.414^{+0.437}_{-0.463}$ from 2D-BAO, $\eta_1=-0.289^{+0.435}_{-0.455}$ from 3D-BAO, and $\eta_{1}=-0.059^{+0.452}_{-0.480}$ from 3D-DESI. 
The constraints on $\eta_{1}$ remain consistent with zero within approximately 1$\sigma$, continuing the trend observed in the previous models.
Moreover, the power-law form of this parameterization provides additional evidence in support of the CDDR and complements the findings from the other parameterizations. }

{For all four parameterizations, the constraints obtained from 2D-BAO, 3D-BAO, and 3D-DESI are consistent and show no significant deviation from $\eta_1 = 0$. These results support the validity of the CDDR. Importantly, this consistency across different types of BAO data suggests that the potential tension between 2D and 3D BAO measurements does not have a significant impact on the parameterized CDDR tests. Nevertheless, it should be noted that the current uncertainties are relatively large, leading to rather weak constraints.} 


{For comparison, the results of other studies using model-independent approaches are presented in Ref.~\cite{2022ApJ...939..115X}, where the CDDR was tested using SNIa data in combination with both low- and high-redshift BAO measurements. In contrast, our analysis combines HII galaxy data with both 2D- and 3D-BAO measurements, including the latest DESI DR2 dataset. This not only extends the redshift coverage, but also allows for a more systematic comparison among different types of BAO data. Compared to Ref.~\cite{2024EPJC...84..702W} and Ref.~\cite{2025EPJC...85..186Y}, who considered only a single type of BAO measurement, our work systematically compares different BAO measurements. Hence, we can assess whether potential tensions between 2D and 3D BAO measurements could affect the parameterized tests of the CDDR. } 

{Our analysis provides a new and independent way to test the CDDR using HII galaxy and BAO data, providing complementary insights that go beyond the conventional SNIa and BAO data framework. The relatively large uncertainties mainly originate from the larger intrinsic scatter of the $L$–$\sigma$ relation compared to the calibrated SNIa luminosities, as well as additional corrections for dust extinction and metallicity that must be marginalized over. Moreover, the current HII galaxy data are smaller in size and have limited signal-to-noise at high redshift, which further broadens the posterior distributions. Nevertheless, the constraints remain centered around $\eta_1 = 0$, showing no strong evidence for deviations from CDDR. This demonstrates that HII galaxy data provide a valuable and independent probe of the CDDR, with the unique potential to extend distance measurements to $z \geq 2$ compared to SNIa.} 


\begin{table*}[htbp]
\centering
\renewcommand{\arraystretch}{2}                                       
\begin{tabular}{|l|c|c|c|c|}
\hline
\textbf{Data} & P1 & P2 & P3 & P4 \\ \hline
{2D-BAO}  &  $-0.054^{+0.220}_{-0.172}$ &  $-0.327^{+0.686}_{-0.448}$  & $-0.166^{+0.409}_{-0.293}$  &  $-0.414^{+0.437}_{-0.463}$ \\
\hline
{3D-BAO} & $0.008^{+0.246}_{-0.162}$  &  $-0.066^{+1.106}_{-0.572}$ &  $-0.004^{+0.598}_{-0.331}$  &  $-0.289^{+0.435}_{-0.455}$ \\ 
\hline
{3D-DESI} &  $0.129^{+0.379}_{-0.214}$ & $-0.047^{+0.584}_{-0.516}$  & $0.037^{+0.344}_{-0.308}$  &  $-0.059^{+0.452}_{-0.480}$ \\ 
\hline
\end{tabular}
\caption{The best-fit values and its 68\% confidence level uncertainties for the parameter $\eta_1$ obtained from the combination of HII galaxy data with 2D-BAO, 3D-BAO, and 3D-DESI BAO measurements, following the procedure described in section~\ref{sec:data}}
\label{tab:results}
\end{table*}
   
\section{Discussions and Conclusions}\label{sec:conclusion}

{Motivated by the existing tension between 2D and 3D BAO measurements, we perform a comprehensive test of the CDDR by combining HII galaxy data with three BAO datasets (2D-BAO, 3D-BAO, and 3D-DESI). This work represents the first robust study of the CDDR using HII galaxies together with BAO data. We aim to quantify whether the existing tension between the angular and anisotropic BAO data would affect the validity of the CDDR. By considering four broad and representative parameterizations of potential CDDR violations, we test the validity of this fundamental relation of cosmology. }

{Our analysis obtains several valuable conclusions: }
\begin{itemize}

    \item {Across the four parameterizations (P1-P4), we find no statistically significant evidence for a violation of the CDDR. The best-fit values of $\eta_1$ from 2D-BAO, 3D-BAO, and 3D-DESI  BAO differ slightly, but their 68\% intervals largely overlap. These results are stable and do not rely on any single functional form. Moreover, the potential tension between 2D and 3D BAO measurements does not significantly affect the constraints of CDDR. This is consistent with the model–independent assessment of Ref.~\cite{2024PhLB..85839027F}. The minor deviation from the standard CDDR cannot be excluded at the present precision, and forthcoming high-precision BAO data will tighten the constraints on this test.    }

    \item {In this work, we present that the HII galaxy data provide an independent perspective on testing the CDDR. Our results provide consistent support for the validity of the CDDR, despite the uncertainties remaining large due to the intrinsic scatter of the $L$–$\sigma$ relation and the limited size of the HII sample. This highlights the importance of exploring alternative distance indicators in cosmological tests.}

\end{itemize}

{Future improvements in HII galaxy and BAO observations, especially at higher redshifts, will provide valuable tools for testing fundamental cosmological principles. It will be crucial to reassess the robustness of CDDR tests as the high-precision BAO data from DESI and the Euclid Space Telescope become available \cite{2012SPIE.8442E..0ZA,2020A&A...644A..80M}. Ideally, both 2D and 3D BAO measurements would be derived from the same collaborations, i.e., DESI \footnote{\url{https://www.desi.lbl.gov/}} and eBOSS \footnote{\url{https://www.sdss4.org/surveys/eboss/}}. It would be particularly useful for determining whether any existing 2D–3D BAO tensions affect the validity of the CDDR. With the forthcoming data from additional and complementary Stage-IV dark energy surveys, such as the Vera C. Rubin Observatory, new opportunities will arise in the next decade. These data will enable tighter constraints on possible deviations from the CDDR and provide insights into its potential redshift evolution. Moreover, exploring alternative parameterizations and combining other cosmological probes will further strengthen our understanding of the CDDR and its implications for fundamental physics.}

\acknowledgments  
We thank the anonymous referee for helpful comments that improved this work. We sincerely thank Tonghua Liu, Fan Yang, and Bharat Ratra for their valuable suggestions and insightful discussions. 
Zheng, J. is supported by the National Natural Science Foundation of China under Grant No. 12403002 and 12433001; The Startup Research Fund of Henan Academy of Sciences (Project No. 241841221); The Scientific and Technological Research Project of Henan Academy of Science (Project No. 20252345001).
Qiang, D.-C. is supported by the National Natural Science Foundation of China under Grant No. 12433001; The Startup Research Fund of Henan Academy of Sciences (Project No. 241841222).
You, Z.-Q. is supported by the National Natural Science Foundation of China under Grant No. 12305059 and 12433001; The Startup Research Fund of Henan Academy of Sciences (Project No. 241841224); The Scientific and Technological Research Project of Henan Academy of Science (Project No. 20252345003); Joint Fund of Henan Province Science and Technology R\&D Program (Project No. 235200810111); Henan Province High-Level Talent Internationalization Cultivation (Project No. 2024032).
Kumar, D. is supported by the Startup Research Fund of the Henan Academy of Sciences under Grant number 241841219.        

   
\bibliographystyle{JHEP}
\bibliography{main.bib}
   

\end{document}